\documentclass[12pt]{article}

\usepackage{amsmath}
\usepackage{amssymb}
\usepackage{graphicx}
\usepackage{cite}
\usepackage{color}
\usepackage{setspace}
\usepackage{caption}
\usepackage{authblk}

\makeatletter
\renewcommand{\@biblabel}[1]{\quad#1.}
\makeatother
\begin{document}

\title{Hierarchy measure for complex networks}
\date{}
\author[1]{Enys Mones}
\author[2]{Lilla Vicsek}
\author[1,3,*]{Tam\'{a}s Vicsek}

\affil[1]{\small{Department of Biological Physics, E\"{o}tv\"{o}s Lor\'{a}nd University, P\'{a}zm\'{a}ny P\'{e}ter stny. 1/A, H-1117 Budapest, Hungary}}
\affil[2]{\small{Institute of Sociology and Social Policy, Corvinus University of Budapest, K\"{o}zrakt\'{a}r u. 4-6., H-1093 Budapest, Hungary}}
\affil[3]{\small{Biological Physics Research Group of HAS, P\'{a}zm\'{a}ny P\'{e}ter stny. 1/A, H-1117 Budapest, Hungary}}

\pagestyle{empty}

\maketitle
\thispagestyle{empty}

\let\oldthefootnote\thefootnote
\renewcommand{\thefootnote}{\fnsymbol{footnote}}
\footnotetext[1]{Corresponding author: \texttt{vicsek@hal.elte.hu}}
\let\thefootnote\oldthefootnote

\begin{abstract}
Nature, technology and society are full of complexity arising from the intricate web of the interactions among the units of the related systems (e.g., proteins, computers, people). Consequently, one of the most successful recent approaches to capturing the fundamental features of the structure and dynamics of complex systems has been the investigation of the networks associated with the above units (nodes) together with their relations (edges). Most complex systems have an inherently hierarchical organization and, correspondingly, the networks behind them also exhibit hierarchical features. Indeed, several papers have been devoted to describing this essential aspect of networks, however, without resulting in a widely accepted, converging concept concerning the quantitative characterization of the level of their hierarchy. Here we develop an approach and propose a quantity (measure) which is simple enough to be widely applicable, reveals a number of universal features of the organization of real-world networks and, as we demonstrate, is capable of capturing the essential features of the structure and the degree of hierarchy in a complex network. The measure we introduce is based on a generalization of the m-reach centrality, which we first extend to directed/partially directed graphs. Then, we define the global reaching centrality (GRC), which is the difference between the maximum and the average value of the generalized reach centralities over the network. We investigate the behavior of the GRC considering both a synthetic model with an adjustable level of hierarchy and real networks. Results for real networks show that our hierarchy measure is related to the controllability of the given system. We also propose a visualization procedure for large complex networks that can be used to obtain an overall qualitative picture about the nature of their hierarchical structure.
\end{abstract}

\section*{Introduction}
The last decade has witnessed an explosive growth of interest in the analysis of complex natural, technological and social systems that permeate many aspects of everyday life. These systems are typically made of many units. Complexity arises from either the structure of the interactions between very similar units or, alternatively, the units and the interactions themselves can have specific characteristics. In both cases, the abstract representation of a complex system can be achieved by a collection of nodes (units) and edges (representing interactions between the units) forming a network (or graph).

Research on networks has considerably profited from using both the standard and novel techniques developed in the field of statistical mechanics \cite{castellano09, vicsek10, pastorsatorras04}. Although a remarkable body of knowledge has accumulated about the statistical properties of networks \cite{barabasi02}, a number of questions are still open. The issue of hierarchy has attracted the attention of a great number of social and natural scientists \cite{pumain06}. It has been argued that hierarchy is present in a wide range of complex systems: such as physical, chemical, biological, and social systems \cite{huseyn84}. Recent empirical findings demonstrate that hierarchy is present in many of the related networks: in the dominant-subordinate hierarchy among animals \cite{goessmann00}, in the hierarchy of the leader-follower network of pigeon flocks \cite{nagy10}, in rhesus macaque kingdoms \cite{fushing11}, in the structure of the transcriptional regulatory network of \emph{Escherichia coli} \cite{ma04}, or in a wide range of social and technological networks \cite{pumain06}. All of these examples suggest that hierarchy is an important feature of natural, artificial and social networks.

It is important to distinguish between the three major types of hierarchies: the \emph{order}, the \emph{nested} and the \emph{flow} hierarchies. In case of an order hierarchy, hierarchy is regarded to be basically only an ``ordered set'', and it is understood to be ``equivalent to an ordering induced by the values of a variable defined on some set of elements'' \cite{lane06} (i.e., generally there is no network behind this concept). In case of a nested hierarchy higher level elements consist of and contain lower level elements, or, as \cite{wimberley09} has formulated ``larger and more complex systems consist of and are dependent upon simpler systems and essential system-component entities''. When a network is structured in a flow hierarchy (mostly directed graphs), the nodes can be layered in different levels so that the nodes that are influenced by a given node (are connected to it through a directed edge) are at lower levels.

Our observation is that the notions of ``hierarchy'' and the ``level of hierarchy'' are very closely related. In fact, without a proper measure of hierarchy the notion of hierarchy cannot be complete. Indeed, there are various definitions of hierarchy, or, in other words, there is no unique, widely accepted definition of the notion of hierarchy itself. Correspondingly, we propose that a good measure of hierarchy can serve as a starting point for finding the best definition of hierarchy.

In this paper, we are interested in flow hierarchy for the following reasons. First, order hierarchy is a single-valued function over the population and there is no underlying network of interactions attached to the hierarchy. Secondly, uncovering a nested hierarchy is analogous to community detection, for which there are known methods \cite{girwan02, palla05}. Finally, both order and nested hierarchies can be converted to flow hierarchies. In an order hierarchy, a directed edge can be assigned to each pair of adjacent members in the hierarchy and this produces a chain of directed edges. In a nested hierarchy, a virtual node is assigned to every subgraph, and if a subgraph contains another, then the two corresponding virtual nodes are connected with a directed link, which produces a flow hierarchy on the network of virtual nodes.

Among the many exciting questions related to hierarchy \cite{pumain06} is concerned with its origin. Several studies have approached this problem from a historical viewpoint \cite{smaje00, dubreuil10} but without any quantitative description. The best known quantitative model for the evolution of hierarchies is the Bonabeau model \cite{theraulaz95}. According to this model, a hierarchy can emerge as the result of the outcomes of competitions between pairs of participating units, and a hierarchy itself is defined by a rank (\emph{order}) assigned to each participating unit \cite{theraulaz95}. Another interesting result comes from game theory: simulations of prisoner’s dilemma type dynamics on adaptive networks showed that cooperation combined with imitation can lead to a hierarchical structure \cite{eguiluz05}. Note, however, that in this model every node can imitate at most one other, and therefore, the emerging hierarchy is by definition a directed tree.

Usually, a hierarchy is the consequence of the different roles, significances and histories of the nodes \cite{theraulaz95, bonabeau99}. In other words, if the influence of the nodes on others (and thence, on the whole system) differs, then a hierarchy can emerge. Nodes with the strongest influence can denote the leaders of a group (as in the structure of a company or hidden groups \cite{rowe07, memon08}; or amongst homing pigeons \cite{nagy10}), central proteins in transcription regulatory networks \cite{ma04, bhardwaj10} or opinion leaders \cite{song07, mak08}. These nodes can have a major impact on the system, and thus, finding them and quantifying the extent of hierarchy at the same time is an important step in the understanding of functionality and controlling of networks.

In most cases networks contain all sorts of edges (both directed and undirected, various edge weights [strength]) making the detection of hierarchy a difficult challenge. When one looks at real-life networks the picture is often much more complicated than for the simple treelike hierarchy: there can be (i) relations between entities on the same level, (ii) ``shortcuts'' when a step in the hierarchy is bypassed, (iii) ties which, instead of going downward on the hierarchy, go upward, (iv) even cycles of connected nodes \cite{hummon95} and (v) clusters \cite{johnsen85}, etc. It can even happen that some or all of the levels of hierarchy cannot be clearly defined (are not well-separated).

The hierarchy measures proposed so far have various undesirable properties that make their application to all classes of complex networks problematic: they (i) use free parameters that are unknown for many networks \cite{rowe07, carmel02}, (ii) quantify only the deviation of the network from the tree and penalize loops or multiple edges \cite{krackhardt94}, and (iii) are applicable only to fully directed or fully undirected graphs \cite{carmel02, rowe07, krackhardt94, trusina04}. Here we are aiming at introducing a measure which can be equally used for all sorts of networks and thus, used for uncovering universal features of the hierarchical organization of the relations within a complex system.

Visualizing the structure of networks has been a widely used approach to obtain a qualitative picture about some of their features (e.g., clusters/modules). At present, the hierarchical visualization of networks is mostly based on the Sugiyama method \cite{sugiyama81}, which offers an informative and clear hierarchical layout for small networks. However, (i) for networks with more than 2-300 nodes the generated layout becomes difficult to understand; (ii) the meaning of the levels is not defined at all; (iii) independently of the presence or absence of a hierarchy in the given network, the method generates a hierarchical layout that is often misleading; (iv) all steps of the Sugiyama method are NP-complete or NP-hard \cite{garey79, healy04}, which makes the usage of several different heuristics necessary and thus, results become less well-defined.

Clearly, there is a need for (a) a measure of hierarchy that is free of the above-mentioned undesired properties and (b) a method for the hierarchical visualization of networks that is unbiased, unambiguous and easily applicable even to large graphs.  Thus, the two main goals of our paper are to provide a universally applicable measure and a visualization technique of the hierarchical structure of complex large networks.

\section*{Definition of the global reaching centrality}
\subsection*{Unweighted directed networks}
We are looking for a measure that is expected to satisfy the following natural and reasonable conditions:
\begin{enumerate}
	\item Absence of free parameters and \emph{a priori} metrics in the definition.
	\item The definition should be for unweighted directed graphs (digraphs) and it should be easily extendable to both weighted and undirected graphs.
	\item The hierarchy measure should be helpful for generating a layout of the graph.
\end{enumerate}

To arrive at an appropriate definition, we quantify the concept of flow hierarchy, where nodes contribute to the dynamics of the system differently. We first define the \emph{local reaching centrality} of node \emph{i} in an unweighted directed graph, \emph{G}, as the generalization of the \emph{m-reach centrality} \cite{borgatti03} to \emph{m = N} (where \emph{N} is the number of nodes in \emph{G}). The local reaching centrality, $C_R(i)$, of node \emph{i} is the proportion of all nodes in the graph that can be reached from node \emph{i} via outgoing edges. In other words, $C_R(i)$ is the number of nodes with a finite positive directed distance from node \emph{i} divided by \emph{N - 1}, i.e., the maximum possible number of nodes reachable from a given node. We aim to define hierarchy as a heterogeneous distribution of the local reaching centrality. Thus, in graph \emph{G} we denote by $C_R^{max}$ the highest local reaching centrality and define the \emph{global reaching centrality} (GRC) as:

\begin{equation}
	GRC=\frac{\sum_{i\in V}\big[C_R^{max}-C_R(i)\big]}{N-1}
	\label{eq:definition_grc}
\end{equation}

Here, \emph{V} denotes the set of nodes in \emph{G}. For normalization, the sum is divided by \emph{N - 1}, as this is the maximal value of the enumerator. In the \emph{GRC = 1} case the graph has only one node with nonzero local reaching centrality (i.e., it is a star graph).

\subsection*{Weighted and undirected networks}
Generalizations to weighted or undirected graphs are straightforward based on the definition of the local reaching centrality. For the generalization of the GRC to weighted directed graphs, we introduce a simple variant of the local reaching centrality:

\begin{equation}
	C'_R(i)=\frac{1}{N-1}\sum_{j:0<d^{out}(i,j)<\infty}\Bigg(\frac{\sum_{k=1}^{d^{out}(i,j)}\omega_i^{(k)}(j)}{d^{out}(i,j)}\Bigg)
	\label{eq:grc_for_weighted}
\end{equation}

Here $d^{out}(i,j)$ is the length of the directed path that goes from \emph{i} to \emph{j} via out-going edges and $\omega_i^{(k)}(j)$ is the weight of the \emph{k}-th edge along this path (link weight is assumed to be proportional to connection strength). If nodes \emph{i} and \emph{j} are connected by more than one directed shortest path, then the one with the maximum weight (i.e., maximum strength) should be used. This extension of the local reaching centrality measures the average weight of a given directed path starting from node i in a weighted directed graph. If we set $\omega_i^{(k)}(j)=1$ for every \emph{i}, \emph{j} and \emph{k}, then the original local reaching centrality (defined for unweighted directed graphs) is recovered.

To generalize the local reaching centrality to undirected unweighted graphs, we remove the $\sum_{k=1}^{d^{out}(i,j)}\omega_i^{(k)}(j)$ term from the previous definition and obtain

\begin{equation}
	C''_R(i)=\frac{1}{N-1}\sum_{j:0<d(i,j)<\infty}\frac{1}{d(i,j)}
	\label{eq:grc_for_undirected}
\end{equation}

This quantity is very similar to the local \emph{closeness centrality} defined by Sabidussi in \cite{sabidussi66}. In fact, this is equivalent to the generalization of the closeness centrality for disconnected graphs given by Opsahl \cite{opsahl10}.

\section*{Methods}
\subsection*{Synthetic model}
\begin{figure}[!ht]
	\begin{center}
	\includegraphics[width=12.35cm]{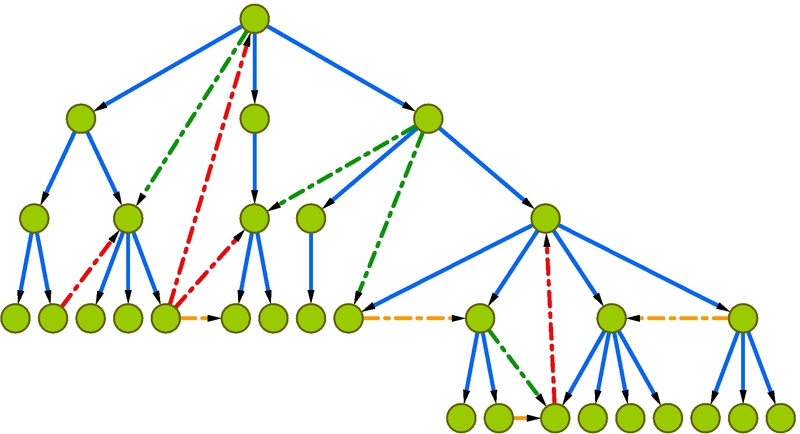}
	\end{center}
	\caption{
	{\bf An adjustable hierarchical network with the different edge types.} The blue edges belong to the original arborescence graph that is used as the backbone of the adjustable hierarchical (AH) network. There are three type of possible edges added to the graph: down edges (green), horizontal edges (orange) and up edges (red). They have different effects on the hierarchical structure of the directed tree. Down edges conserve the hierarchy, horizontal edges has a slight influence and up edges make strong changes in the structure.
}
	\label{fig:fig1}
\end{figure}

In order to show the behavior of GRC, we introduce a synthetic network model with tunable extent of hierarchy. The construction of the network is the following:
\begin{enumerate}
	\item In a directed tree assign a level ($\ell$) to every node. The level of the root node is equal to the number of levels. If and only if a node has level $\ell$, then the level of its children will be $\ell-1$. These levels denote the natural layers in the hierarchy of the directed tree (the nodes at the bottom have $\ell=1$).
	\item We put a given number of additional random directed edges in the graph according to the following rule. \emph{1 - p} proportion of the edges is totally random, i.e. we choose two nodes randomly (\emph{A} and \emph{B}) and if they are not already connected in the given ($A\to B$) direction, we connect them. By \emph{p} proportion of the edges, we put the $A\to B$ edge only if $\ell_A>\ell_B$. In this way, \emph{p} proportion of the random edges will not change the hierarchical structure of the directed tree.
\end{enumerate}
An example of a generated network with the different edge types is shown in Figure \ref{fig:fig1}. Hereafter, we will refer to this synthetic model as the \emph{adjustable hierarchical network} (AH). 

\subsection*{Randomization of real networks}
During the analysis of the results with real networks, we also calculated the GRC after randomizing them: first, we generated a random network with the same in and out degree distribution according to the configuration model. The generated network is further randomized in the following way: we choose two random edges ($A\to B$ and $C\to D$) and change the endpoints of them (so that we get $A\to D$ and $C\to B$). In every case, the number of rewired edge pairs was ten times the number of edges.

\subsection*{Visualization}
We also propose a visualization method using an arbitrary local quantity on the graph. The algorithm is as follows:
\begin{enumerate}
	\item Grade the nodes according to the local quantity $x_i$.
	\item Add nodes to the first (lowermost) level of the layout in the increasing order of their $x_i$ values as long as $\sigma_L<z\cdot\sigma_G$. Here $\sigma_L$ is the standard deviation of $x_i$ within the current (first) level, $\sigma_G$ is the standard deviations of $x_i$ within the whole graph, and \emph{z} is an adjustable coefficient.
	\item When $\sigma_L\ge z\cdot\sigma_G$ is reached, start a new level.
	\item Repeat 2nd and 3rd steps until every node is put in levels.
	\item For horizontal arrangement, align the center of every level to the same vertical line. In other words, in each level, the average of the horizontal positions of the nodes is the same:
\begin{equation*}
	X_{\ell_1}=X_{\ell_2}=0\qquad\textrm{for all $\ell_1$ and $\ell_2$}
\end{equation*}
Here, $X_\ell$ is the horizontal center of mass of level $\ell$.
	\item The levels are arranged vertically so that the distances between adjacent levels are proportional to the logarithm of the differences in the averages inside the corresponding levels, i.e.
\begin{equation*}
	(Y_{\ell+1}-Y_{\ell})\quad\propto\quad\ln\big[\langle x\rangle_{\ell+1}-\langle x\rangle_{\ell}\big]
\end{equation*}
where $Y_{\ell}$ and is the vertical position of the $\ell$-th level and $\langle x\rangle_{\ell}$ is the average of $x_i$ inside this level. First, set the vertical distances of levels proportionally to the differences between their average values of $x_i$ such that the smallest distance will be set to a given length (this length is the same as the horizontal distance between two adjacent nodes). Finally, set the distances to be proportional to the logarithm of the original differences so that the height of the graph is kept unchanged.
\end{enumerate}
In the above steps we use the standard deviation in order to get clearly different layouts for different distributions of $x_i$. In a network with a localized distribution of $x_i$ the method produces few levels that are very close to each other. But if the distribution of $x_i$ is non-localized, the network will have many levels and a large vertical extension. If the distribution of $x_i$ is continuous, then we can use \emph{z} to adjust the extent to which every level contributes to the total variance. In other words, for large graphs, \emph{z} tunes the vertical extension of the layout. If the distribution of $x_i$ is discrete, then we can assign a level to each of its different values, which is mathematically equivalent to \emph{z = 0}. In practice, we set \emph{z} to a sufficiently small value, $\varepsilon$.

For the graph generations, randomizations and shortest path calculations, we used the already implemented functions in the \emph{igraph} software package \cite{csardi06}.

\section*{Results}
\subsection*{Classical random networks}
\begin{figure}[!ht]
	\begin{center}
	\includegraphics[width=12.35cm]{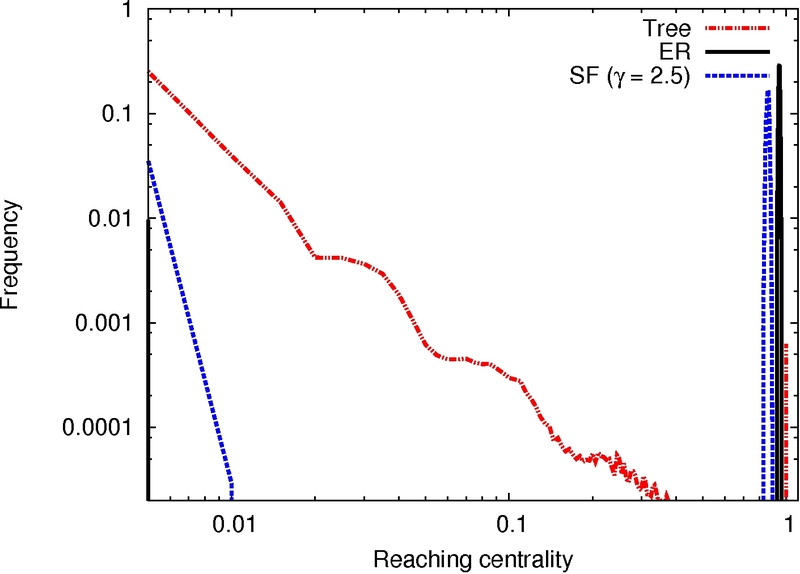}
	\end{center}
	\caption{
	{\bf Distributions of the local reaching centrality for different network types.} For each network type $N=2000$ and for the Erd\H{o}s--R\'{e}nyi (ER) and scale-free (SF) networks $\langle k\rangle=3$. All curves show averages of the distributions over an ensemble of 1000 graphs. Standard deviations are comparable with the averages only near the peaks in the ER and SF models. Although the standard deviations at the peaks are large, they do not change the positions of the peaks, and thus, do not affect the distributions.
}
	\label{fig:fig2}
\end{figure}

In order to demonstrate the basic features of the GRC, we briefly discuss its behavior for a few well-known network types. For Erd\H{o}s--R\'{e}nyi (ER) graphs \cite{erdos60, bollobas01}, scale-free (SF) \cite{barabasi99, goh01, chung02} graphs and directed trees (more precisely arborescences with random branching number \cite{tutte01, grinstead97}), the distribution of $C_R$ is markedly different (the curves in Figure \ref{fig:fig2} are averages for 1000 random graphs of each type). In every case, the exponent for the SF networks was set to $\gamma=2.5$. For the directed tree, the distribution follows a power-law that is distorted due to the random branching numbers. Directed trees have a maximally heterogeneous distribution of $C_R$, thus, based on our arguments above, they are maximally hierarchical. Note that the hierarchical tree (directed tree) has very few nodes with local reaching centrality close to 1.
This is in contrast with the ER and SF graphs in which most of the nodes have a large local reaching centrality. Since almost every node has the same centrality, the contribution of the nodes in Eq. \ref{eq:definition_grc} for the ER and SF graphs is negligible. Note that not only the GRC, but also the standard deviation of $C_R$ increases with the heterogeneity of the graph. The values of GRC are shown in Table 1 together with the standard deviation of the distribution. However, the GRC itself is more suitable for quantifying the heterogeneity of the graph for two reasons. On the one hand, the accuracy of the standard deviation of $C_R$ is worse than that of the GRC (it has larger deviation on the ensemble of graphs). On the other hand, the standard deviation of $C_R$ is much smaller for the directed tree than for the ER, which is in contrast to our definition making the tree maximally hierarchical. In summary, we find that, based on their reaching centralities, ER graphs are not hierarchical at all, as expected, and SF graphs are slightly hierarchical.
\\
\begin{table}[!ht]
	\centering
	\caption{
	\bf{Heterogeneity of the distribution of the local reaching centrality for different network types.}}
	\begin{tabular}{|c|c|c|}
		\hline
		\textbf{Graph} & $\mathbf{GRC}$ & $\mathbf{\sigma(C_R)}$ \\
		\hline
		ER & $0.058\pm0.005$ & $0.222\pm0.010$ \\
		\hline
		SF & $0.127\pm0.008$ & $0.300\pm0.009$ \\
		\hline
		Tree & $0.997\pm0.001$ & $0.031\pm0.004$ \\
		\hline
	\end{tabular}
	\caption*{The two measures of heterogeneity presented here are the global reaching centrality ($GRC$) and $\sigma(C_R)$ (standard deviation of $C_R$). Means and variances are shown for an ensemble of 1000 networks.}
	\label{tab:table1}
\end{table}

\subsection*{Adjustable hierarchical network}
\begin{figure}[!ht]
	\begin{center}
	\includegraphics[width=12.35cm]{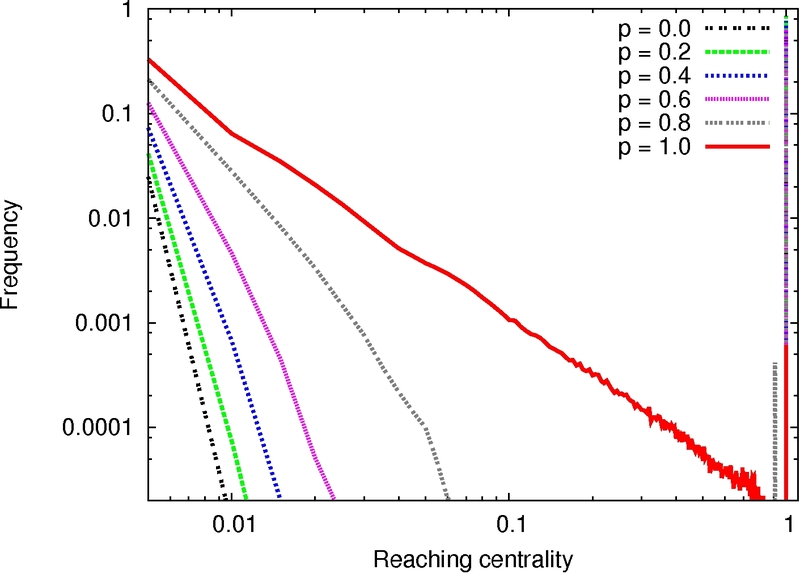}
	\end{center}
	\caption{
	{\bf Distribution of the local reaching centrality for the adjustable hierarchical network.} Distribution of the local reaching centrality in the adjustable hierarchical (AH) network model at different $p$ parameter values. Each distribution is averaged over 1000 AH networks with $N=2000$ and $\langle k\rangle=3$. The standard deviations of the distributions are comparable to the averages only for relative frequencies less than 0.002. Note that from the $p=0$ (highly random) to the $p=1$ (fully hierarchical) state the distribution changes continuously and monotonously with $p$.
}
	\label{fig:fig3}
\end{figure}

\begin{figure}[!ht]
	\begin{center}
	\includegraphics[width=12.35cm]{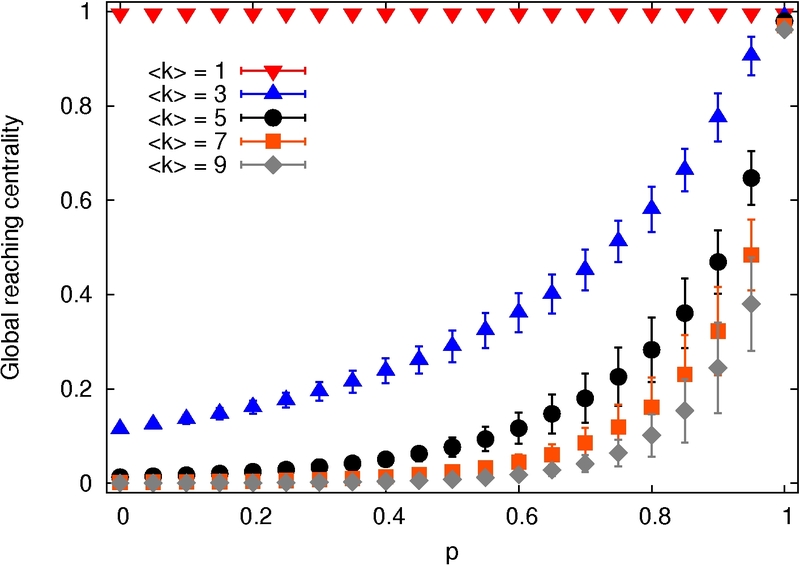}
	\end{center}
	\caption{
	{\bf The global reaching centrality at different p values in the adjustable hierarchical model.} All curves show averages over an ensemble of 1000 networks with $N=2000$ and different average degrees. Standard deviations grow with $p$, but they are clearly below the average values of the GRC. Note that for larger density, it is less likely to obtain the same level of hierarchy.
}
	\label{fig:fig4}
\end{figure}

\begin{figure}[!ht]
	\begin{center}
	\includegraphics[width=12.35cm]{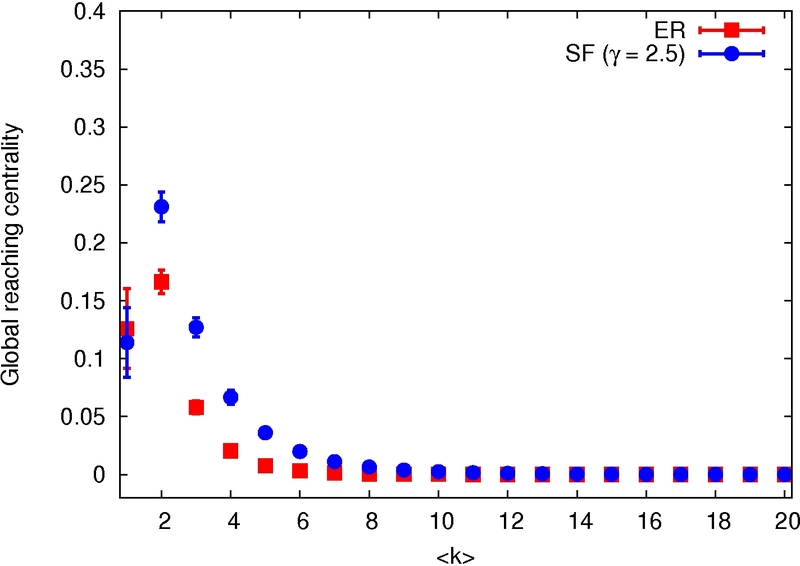}
	\end{center}
	\caption{
	{\bf The global reaching centrality versus average degree in the Erd\H{o}s--R\'{e}nyi and scale-free networks.} Dots show averages for 1000 graphs with $N=2000$ nodes. In the Erd\H{o}s--R\'{e}nyi and scale-free networks, standard deviations of the GRC are comparable with its averages only for $\langle k\rangle>7$ and $\langle k\rangle>12$, respectively.
}
	\label{fig:fig5}
\end{figure}

In the $p=0$ limit, the topology of the AH graph is close to that of an ER graph, but, as one can see, the distribution of the local reaching centrality values of the AH is similar to that of the SF network (Figure \ref{fig:fig3}): a little wider at small centralities than in the ER case. By increasing p, the distribution further widens around the origin and at \emph{p = 1}, it resembles the one for the directed tree, but it is even closer to a power-law. The global reaching centrality as function of the parameter \emph{p} is shown in Figure \ref{fig:fig4}. The GRC monotonously increases with \emph{p} and sweeps through the (0,1) interval in the synthetic model, indicating that it is suitable for measuring the level of hierarchy. As seen in the figures, the global reaching centrality at a given value of \emph{p} is less for larger average degrees. This observation is confirmed with the results on ER and SF networks (Figure \ref{fig:fig5}). For large densities the GRC vanishes for both the ER and the SF networks.

\clearpage
\subsection*{Real networks}
We now turn our attention to the hierarchical properties of real networks. The global reaching centralities for different types of networks are shown in Table 2. For each network we show the average degree ($\langle k\rangle$) and the GRC of the real network. It is important to point out that the direction of the edges in real networks had to be well-defined before calculating the reaching centrality. In every case, the networks were directed so that the source of an edge had a larger effect on the target than conversely. This choice of directedness originates in the observation that the higher a node is in the hierarchy, the more impact it has on the network. According to Table 2, the GRC can have values from a broad range, depending on the average degree and the structure of the networks. For graphs with higher average degree, the GRC is usually smaller. This indicates that for a dense network it is harder to achieve a large reaching centrality, as seen with the ER, SF and AH graphs. The value of the GRC shows how hierarchical the structure of the network is. Food webs have the largest GRC and networks of intra-organizational trust have the smallest. This is in good agreement with the extremely low number of loops in food webs and the high number of loops in email-based organizational networks.

\begin{table}[!ht]\footnotesize
	\centering
	\caption{
	\bf{Hierarchical properties of real networks.}}
	\begin{tabular}{|c|c|c|c|c|c|c|}
	\hline
	\textbf{Type}	& \textbf{Meaning of} $\mathbf{A\to B}$	& \textbf{Network}	& $\mathbf{N}$ & $\mathbf{\langle k\rangle}$& $\mathbf{GRC}$& $\mathbf{GRC^{rand}}$ \\
	\hline
	Food web	& A eats B			& \textcolor{red}{Ythan} \cite{dunne02}		& 135	& 4.452	& 0.814		& 0.507 \\
	\hline
			&				& \textcolor{red}{Seagrass} \cite{christian99}	& 49	& 4.612	& 0.723		& 0.253 \\
	\hline
			&				& \textcolor{red}{LittleRock} \cite{martinez91}	& 183	& 13.628 & 0.811	& 0.045 \\
	\hline
			&				& \textcolor{red}{GrassLand} \cite{dunne02}	& 88	& 1.557	& 0.961		& 0.695 \\
	\hline
	Electric	& B depends on the value at A	& \textcolor{red}{s1488} \cite{circuits}	& 667	& 2.085	& 0.482		& 0.298 \\
	\hline
			&				& \textcolor{red}{s1494} \cite{circuits}	& 661	& 2.116	& 0.482		& 0.289	\\
	\hline
			&				& \textcolor{red}{s5378} \cite{circuits}	& 2993	& 1.467	& 0.231		& 0.062	\\
	\hline
			&				& \textcolor{red}{s9234} \cite{circuits}	& 5844	& 1.4	& 0.424		& 0.05 \\
	\hline
			&				& \textcolor{red}{s35932} \cite{circuits}	& 17828	& 1.683	& 0.459		& 0.015 \\
	\hline
	Metabolic	& B is an end product of A	& \emph{C. elegans} \cite{jeong00}			& 1173	& 2.442	& 0.048		& 0.052 \\
	\hline
			&				& \textcolor{red}{\emph{E. coli}} \cite{jeong00}	& 2275	& 2.533	& 0.043		& 0.058 \\
	\hline
			&				& \emph{S. cerevisiae} \cite{jeong00}			& 1511	& 2.537	& 0.037		& 0.042 \\
	\hline
	Neuronal	& A synapse goes from A to B	& \textcolor{red}{\emph{C. elegans}} \cite{achacoso92, watts98}	& 297	& 7.943	& 0.133		& 0.023 \\
	\hline
			&				& Macaque brain \cite{negyessy06}		& 45	& 10.289 & 0		& 0 \\
	\hline
	Internet	& A communicates with B 	& p2p-1 \cite{leskovec05, ripeanu02}		& 10876	& 3.677	& 0.598		& 0.597 \\
	\hline
			&				& p2p-2 \cite{leskovec05, ripeanu02}		& 8846	& 3.599	& 0.6		& 0.599 \\
	\hline
			&				& p2p-3 \cite{leskovec05, ripeanu02}		& 8717	& 3.616	& 0.607		& 0.605 \\
	\hline
	Organization	& B trusts in A			& \textcolor{blue}{Enron} \cite{leskovec09, klimmt04}	& 156	& 10.699 & 0.038	& 0.044 \\
	\hline
			&				& \textcolor{red}{Consulting} \cite{cross04}	& 46	& 19.109 & 0.043	& 0.032 \\
	\hline
			&				& Manufacturing \cite{cross04}			& 34	& 18.935 & 0.013	& 0.013 \\
	\hline
			& B knows A			& \textcolor{blue}{Freemans-1} \cite{freeman79}	& 34	& 18.971 & 0.028	& 0.041 \\
	\hline
			&				& Freemans-2 \cite{freeman79}			& 77	& 24.412 & 0		& 0 \\
	\hline
	Trust		& B trusts in A			& \textcolor{blue}{WikiVote} \cite{leskovec10}	& 7115	& 14.573 & 0.494	& 0.534 \\
	\hline
			&				& College \cite{vanduijn03, milo04}		& 32	& 3	& 0.275		& 0.273 \\
	\hline
			&				& \textcolor{red}{Prison} \cite{vanduijn03, milo04}	& 67	& 2.716	& 0.172		& 0.111 \\
	\hline
	Language	& B follows A			& \textcolor{blue}{English} \cite{cancho01}	& 7724	& 5.992	& 0.128		& 0.238 \\
	\hline
			&				&\textcolor{blue}{French} \cite{cancho01}	& 9424	& 2.578	& 0.657		& 0.875 \\
	\hline
			&				& Spanish \cite{cancho01}			& 12642	& 3.57	& 0.951		& 0.939	\\
	\hline
			&				& \textcolor{blue}{Japanese} \cite{cancho01}	& 3177	& 2.613	& 0.054		& 0.206 \\
	\hline
	Regulatory	& A regulates B			& \textcolor{blue}{TRN-Yeast-1}	\cite{balaji06}	& 4441	& 2.899	& 0.934		& 0.968 \\
	\hline
			&				& \textcolor{blue}{TRN-Yeast-2}	\cite{milo02}	& 688	& 1.568	& 0.116		& 0.67 \\
	\hline
			&				& \textcolor{blue}{TRN-EC} \cite{milo02}	& 419	& 1.239	& 0.261		& 0.679 \\
	\hline
	\end{tabular}
	\caption*{We show the order ($N$), average degree ($\langle k\rangle$), and global reaching centrality for the original ($GRC$) and for the randomized networks ($GRC^{rand}$). References to data sources are included. Networks shown in red are the most hierarchical (compared to their randomized versions) and networks in blue are more egalitarian (with a 98\% confidence interval). The meaning of edges is also indicated.}
	\label{tab:table2}
\end{table}

While the actual value of the GRC provides information about the hierarchical properties of the network, we can also compare the results to the randomized versions of the original networks to see how consistent the value we obtained is with the expectations. In order to do this, for each network we generated 100 random networks with the same degree (the details of randomization is explained in the Methods section): the mean values of the global reaching centralities for these randomized networks are shown in Table 2 ($GRC^{rand}$). The color of the networks' names indicates the relation of each original network to its randomized version: the names of statistically significantly (with a confidence interval of 98\%) hierarchical networks are in red while the names of non-hierarchical ones (same confidence) are in blue. Apart from the actual GRC values, the comparison to randomized networks by $GRC/GRC^{rand}$ shows slight differences between the analyzed network types. For the food webs $GRC/GRC^{rand}$ is remarkably high. Although the electronic circuits have low GRC values, they are significantly more hierarchical than their randomized versions. In contrast, although the Internet networks have larger reaching centralities than most other listed networks, these values do not differ significantly from the values of the corresponding randomized networks. Also note that the regulatory networks are significantly less hierarchical, mostly because biochemical systems contain many feedbacks keeping the processes stabilized.

The emergence of hierarchy in many human-made organizations and networks raises the question whether conscious control over these systems plays a role in the origin of hierarchy? In order to investigate this question, we compared the global reaching centralities with the controllability of networks as defined by Liu et al. \cite{liu11}. They show that the minimal number of \emph{driver nodes} ($N_D$) is related to the maximum matching of the network and they also provide an algorithm for determining $N_D$. In a network with \emph{N} nodes the relative number of driver nodes is $n_D^{Liu}=N_D/N$. Driver nodes are the nodes that have to be controlled in order to take full control over the network. Full control means that one can drive the system from any initial state to any other desired final state. Since the networks listed in Table 2 have different original functions (food web, electric, etc.), and in many cases their controllability and hierarchical properties are not yet well understood, we compared these two quantities separately within each group of networks. The Pearson correlations of the GRC and $n_D^{Liu}$ are shown in Table 3. In most of the listed real networks, the correlation is above 0.5, which is a relatively small value but still indicates a weak relation between the two quantities. Next, we compared the hierarchy measure, GRC, to the ratio of driver nodes in our synthetic model. Interestingly, for high link densities ($\langle k\rangle\ge5$) the ratio of driver nodes is very close to the value of the GRC and they differ significantly only for highly hierarchical graphs (i.e., for $p>0.85$). In an easily (hardly) controllable network, i.e., where $n_D$ is low (high), few (many) nodes need to be controlled for a total control over the network. According to the results shown in Table 3 for real graphs and the results with the synthetic model (for a wide range of \emph{p}) the GRC and  are moderately positively correlated. In other words, \emph{hierarchical networks are harder to control}. This result contradicts our initial intuitive concept that hierarchy emerges because it is the optimal structure with respect to controllability. This contradiction can be traced back to an assumption in the node-based definition of controllability given in \cite{liu11} where each node is assumed to send the same signal to all of its neighbors. If, however, the network's dynamics is defined on the edges \cite{nepusz11}, then the definition of controllability differs from the definition by Liu et al. Therefore, as an alternative, we compared hierarchy to controllability defined under the \emph{switchboard dynamics} \cite{nepusz11} (correlations are shown in Table 4). In the case of switchboard dynamics edges are controlled and nodes are simple devices converting the signals arriving on their in-edges to signals leaving on their out-edges. The driver nodes in this dynamics are those that one has to control for controlling the state of every edge. Based on the correlations between the GRC and the number of driver nodes, we conclude that \emph{under the switchboard dynamics hierarchical networks are better controllable}.

\begin{table}[!ht]
	\centering
	\caption{
	\bf{The Pearson correlation of the GRC and $n_D$ defined by Liu et al.}}
	\begin{tabular}{|c|c|}
	\hline
	\textbf{Type of the networks}	& $\mathbf{\rho(GRC,n_D^{Liu})}$ \\
	\hline
	Regulatory			& 0.843 \\
	\hline
	Trust				& 0.974 \\
	\hline
	Food web			& 0.69 \\
	\hline
	Metabolic			& -0.225 \\
	\hline
	Electric			& 0.503 \\
	\hline
	Internet			& 0.632 \\
	\hline
	Organizational			& 0.337 \\
	\hline
	Language			& 0.933 \\
	\hline
	\end{tabular}
	\caption*{With only one exception, all correlations are positive and many of them are above 0.6, i.e., the GRC and $n_D^{Liu}$ are positively correlated.}
	\label{tab:table3}
\end{table}

\begin{table}[!ht]
	\centering
	\caption{
	\bf{Pearson correlation of the GRC and $n_D$ in the switchboard dynamics.}}
\begin{tabular}{|c|c|}
	\hline
	\textbf{Type of the networks}	& $\mathbf{\rho(C_R,n^{SBD}_D)}$ \\
	\hline
	Regulatory			& -0.922 \\
	\hline
	Trust				& -0.983 \\
	\hline
	Food web			& -0.406 \\
	\hline
	Metabolic			& -0.916 \\
	\hline
	Electric			& -0.969 \\
	\hline
	Internet			& 0.57 \\
	\hline
	Organizational			& -0.674 \\
	\hline
	Language			& -0.812 \\
	\hline
	\end{tabular}
	\caption*{The correlations are all negative (except for the Internet networks) and most of them are very close to -1. Thus, under the switchboard dynamics the GRC (strength of hierarchy) and $n_D^{SBD}$ are strongly negatively correlated.}
	\label{tab:table4}
\end{table}

To show how the generalized reaching centralities can be applied to undirected networks, we tested our method on the networks of terrorists investigated by Memon et al. Our results are similar to those of \cite{memon08}: the top of the hierarchy related to the Bojinka case contains Isamudin and K. S. Mehmood (known as Mohammed). In the London Bombings network \cite{memon08} found  that the mastermind of the 7/7 bombings was H. R. Awsat; he was identified by our analysis (based on $C''_R$) as a leader and M. S. Khan and I. M. Said as additional important participants. These results suggest that the above extensions of the local reaching centrality are effective quantities for the description of undirected graphs.

\subsection*{Visualization of large networks}
\begin{figure}[!ht]
	\begin{center}
	\includegraphics[width=12.35cm]{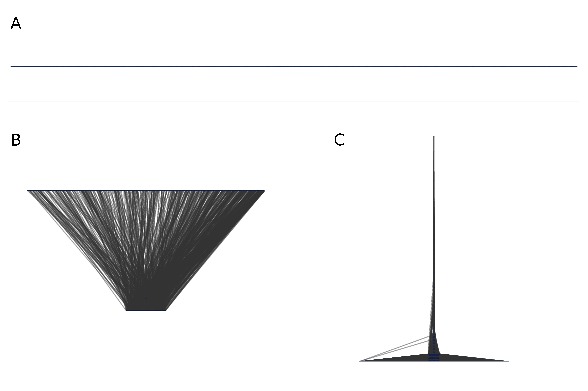}
	\end{center}
	\caption{
	{\bf Visualization of three network types based on the local reaching centrality.} Visualization of (A) an Erd\H{o}s--R\'{e}nyi (ER) network, (B) a scale-free (SF) network and (C) a directed tree with random branching number between 1 and 5. All three graphs have $N=1000$ nodes and the ER and SF graphs have $\langle k\rangle=3$. In each network $z$ was set to $2/N$.
}
	\label{fig:fig6}
\end{figure}

We use the method introduced in the Methods section for the hierarchical visualization of unweighted digraph by setting $x_i=C_R(i)$. Since the local reaching centrality takes discrete values on the graph, we use $z=\varepsilon$, that is, nodes that have local reaching centralities very close to each other are in the same level. Figure \ref{fig:fig6} shows the layout of various graphs. ER graphs have only two layers close to each other and most of their nodes are in the top layer indicating an almost equal impact of every node and the absence of hierarchy. As opposed to this, an arborescence has many layers, the distances between the layers vary and the layers contain different numbers of nodes. At the topmost layer there is only one node and it is far from the other nodes. This structure is due to the fact that the roles of nodes in the graph vary on a wide range, in other words, the distribution of the local reaching centrality is strongly heterogeneous. The hierarchical structure of an SF graph is between those of an ER graph and an arborescence: although it has only a few layers, these layers are clearly separated.

\begin{figure}[!ht]
	\begin{center}
	\includegraphics[width=12.35cm]{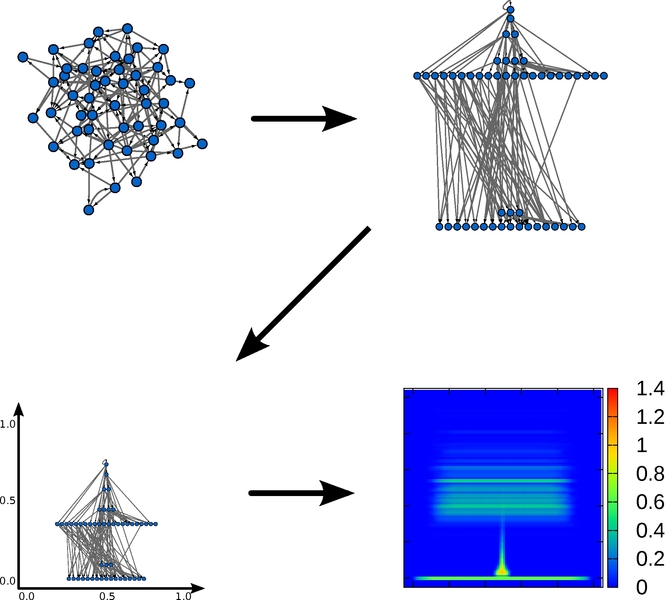}
	\end{center}
	\caption{
	{\bf Diagram illustrating the process of visualizing an ensemble of networks.} First, we compute the layout based on the selected $x_i$ local quantity for each graph in the ensemble (top right). Next, we separate the levels logarithmically and scale each layout into the unit square (bottom left). Last, we overlay all rescaled layouts and plot the obtained density of nodes in the unit square (bottom right, see color scale also). In the heat maps, the color scale shows $\log(\log(\rho(x,y)+1)+1)$, where $\rho(x,y)$ is the average density of the ensemble.
}
	\label{fig:fig7}
\end{figure}

\begin{figure}[!ht]
	\begin{center}
	\includegraphics[height=17.35cm]{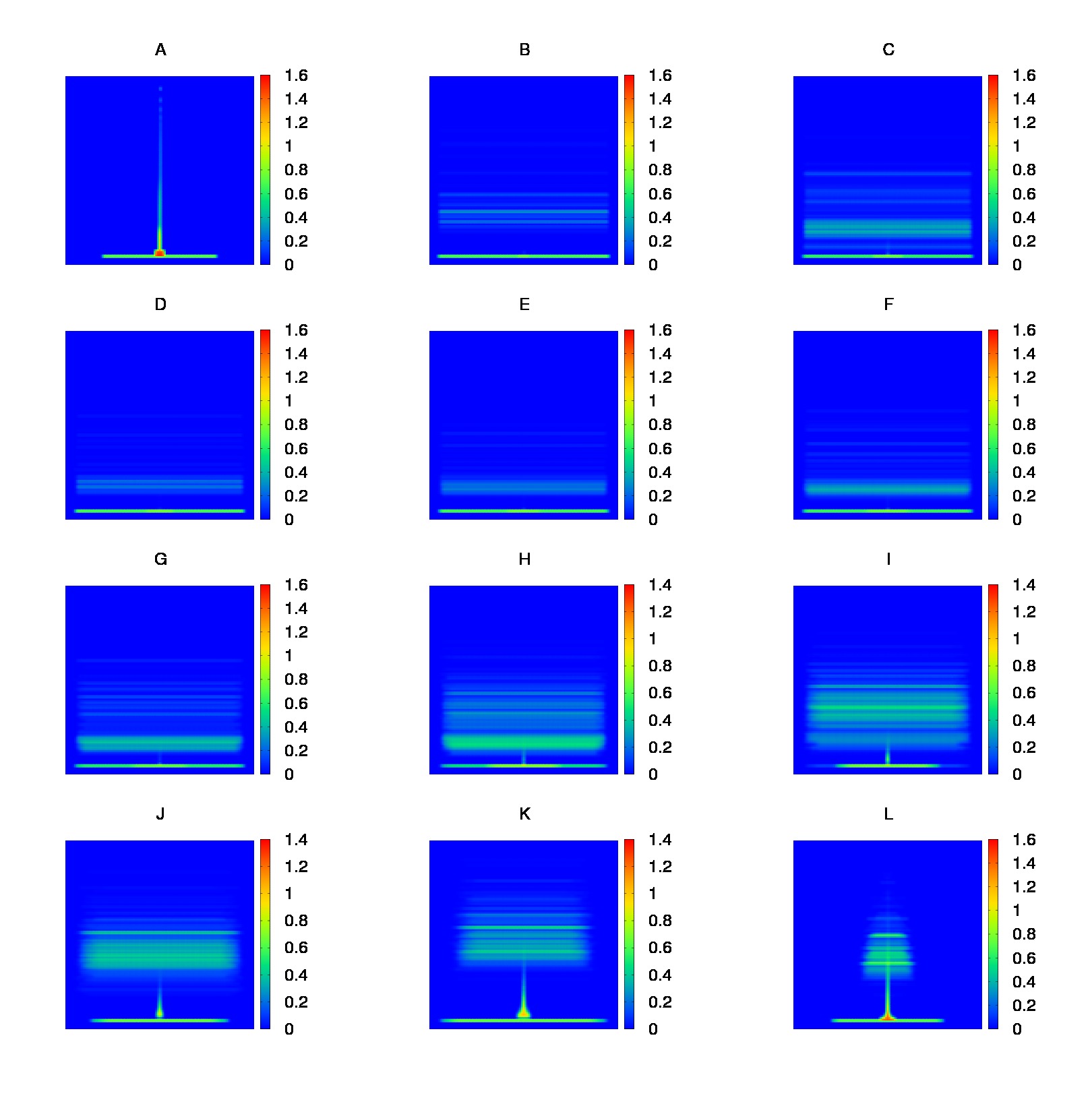}
	\end{center}
	\caption{
	{\bf Visualization of network ensembles.} Visualizations of the (A) Erd\H{o}s--R\'{e}nyi, (B) scale-free, (C) directed tree and (D)-(L) AH network ensembles (subfigures (D)-(L) are for different values of the model parameter: $p=0.1,\dots,0.9$). In each case the color scale shows $\log(\log(\rho(x,y)+1)+1)$ where $\rho(x,y)$ is the density averaged over 1000 graphs. $N=2000$ and $\langle k\rangle=3$ were set. In every network, $z$ was set to $3/N$. The corresponding GRC values are: 0.997 (A), 0.058 (B), 0.127 (C), 0.135 (D), 0.161 (E), 0.194 (F), 0.238 (G), 0.290 (H), 0.361 (I), 0.452 (J), 0.581 (K) and 0.775 (L).
}
	\label{fig:fig8}
\end{figure}

\begin{figure}[!ht]
	\begin{center}
	\includegraphics[width=12.35cm]{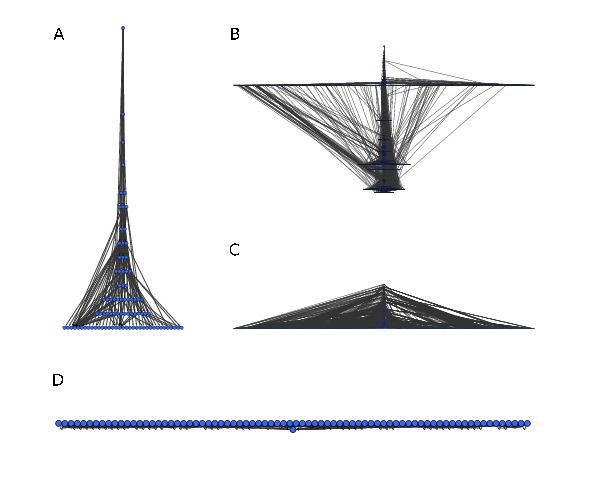}
	\end{center}
	\caption{
	{\bf Visualization of real networks.} The hierarchy-based visualization of (A) the GrassLand food web, (B) the electrical circuit benchmark s9234, (C) the transcriptional regulatory network of yeast  and (D) the core of the Enron network. In every network $z$ was set to $2/N$.
}
	\label{fig:fig9}
\end{figure}

Note that different realizations (single graphs) of the same graph model (e.g., the SF model) usually have different hierarchical layouts. In order to eliminate this bias and to compare the graph models themselves (instead of single graphs from each model), we apply the hierarchical layouts of single graphs to define the drawing (image) of graph ensembles. To do this, first we rescale the hierarchical layout of each single graph to unit height and width and center it in the unit square (Figure \ref{fig:fig7}). Next, we overlay the hierarchical layouts of graphs from the same model. For each graph model the result of this process is a density distribution of the nodes (in the unit square) averaged over the different realizations of the given model. Figure \ref{fig:fig8} shows graph ensemble drawings: the ER model is visualized as a thin horizontal line at the bottom of the box, while the SF model has more levels and it is similar to the AH(0.3) network. The ensemble of arborescences is visualized in a small concentrated region at the bottom of the unit square indicating the presence of many close levels. The transition from egalitarianism to hierarchy can be clearly seen on the visualization of the AH graphs. At small \emph{p} (proportion of edges pointing to a lower level) there is mostly one level, then with increasing \emph{p} more and more other levels emerge, and finally, the network splits into two groups of levels that are moving away from each other. To illustrate the usefulness of our visualization method, we show results for four real graphs as well (Figure \ref{fig:fig9}). The GrassLand network is highly hierarchical, while the Enron network is very egalitarian (only very few nodes are much lower than the majority). This is in good agreement with the global reaching centrality values. The electrical circuit and the biological regulatory network are between the two extreme cases. The first contains two major levels (further subdivided into smaller levels. In contrast, the regulatory network has only one wide bottom level and a few nodes in the top and they are close to each other.

\clearpage
\section*{Discussion}
Hierarchy is an essential feature of many natural and human-made networks and therefore, it is of high importance to have a measure quantifying it.  Here we proposed a measure based on the assumption that the rank of the nodes should be related to their impact on the whole network, which is proportional to the number of all nodes reachable from them (local reaching centrality). The quantity we introduced, i.e., the global reaching centrality (GRC), measures the heterogeneity of the local reaching centrality distribution on the whole graph. In contrast to formerly proposed measures, the GRC does not penalize loops and undirected edges, but takes them into account by making bidirectionally connected pairs of nodes ($A\to B$, $B\to A$) equivalent in the hierarchy. There are neither free parameters in the method, nor optimization, and the ranks of the nodes are a natural result of the GRC. Since the controllability (according to the switchboard dynamics) and the extent of hierarchy are positively correlated, our calculations indicated that hierarchical structures are more easily controllable.

\section*{Acknowledgements}
We thank Ill\'{e}s Farkas and G\'{a}bor V\'{a}s\'{a}rhelyi for their helpful comments on the early version of the manuscript. We also thank Tam\'{a}s Nepusz for his technical and theoretical advices and suggestions on the simulations.

\bibliography{HierarchyMeasureForComplexNetworks}

\end{document}